\documentclass[12pt,a4paper]{article}
\usepackage[utf8]{inputenc}
\usepackage{amsmath}
\usepackage{amssymb}
\usepackage{graphicx}
\usepackage{booktabs}
\usepackage{longtable}
\usepackage{amssymb}
\usepackage{array}
\usepackage{subcaption}
\usepackage{hyperref}
\hypersetup{
    colorlinks=true, 
    linkcolor=black, 
    citecolor=black, 
    filecolor=black, 
    urlcolor=black   
}
\usepackage{geometry}
\usepackage[
backend=biber,
style=chem-angew,
doi=false,
isbn=false,
url=false,
eprint=false
]{biblatex}

\newcolumntype{C}[1]{>{\centering\arraybackslash}p{#1}}

\title{A potassium ion channel simulated with a universal neural network potential.}

\author{Tim Duignan \\ 
\texttt{tim@orbitalmaterials.com} \\
Orbital Materials \\
School of Chemical Engineering, The University of Queensland}

\date{\today}

\begin{document}

\maketitle
\begin{abstract}
Potassium ion channels are critical components of biology. They conduct potassium ions across the cell membrane with remarkable speed and selectivity. Understanding how they do this is crucially important for applications in neuroscience, medicine, and materials science. However, many fundamental questions about the mechanism they use remain unresolved, partly because it is extremely difficult to computationally model due to the scale and complexity of the necessary simulations. Here, the selectivity filter (SF) of the KcsA potassium ion channel is simulated using Orb-D3, a recently released universal neural network potential. A previously unreported hydrogen bond between water in the SF and the T75 hydroxyl side group at the entrance to the SF is observed. This hydrogen bond appears to stabilize water in the SF, enabling a soft knock-on transport mechanism where water is co-transported through the SF with a reasonable conductivity (80 $\pm$ 20 pS).  Carbonyl backbone flipping is also observed at new sites in the SF. This work demonstrates the potential of universal neural network potentials to provide insights into previously intractable questions about complex systems far outside their training data distribution.
\end{abstract}
\maketitle

\begin{figure}[tbh]
\centering
\includegraphics[width=\columnwidth]{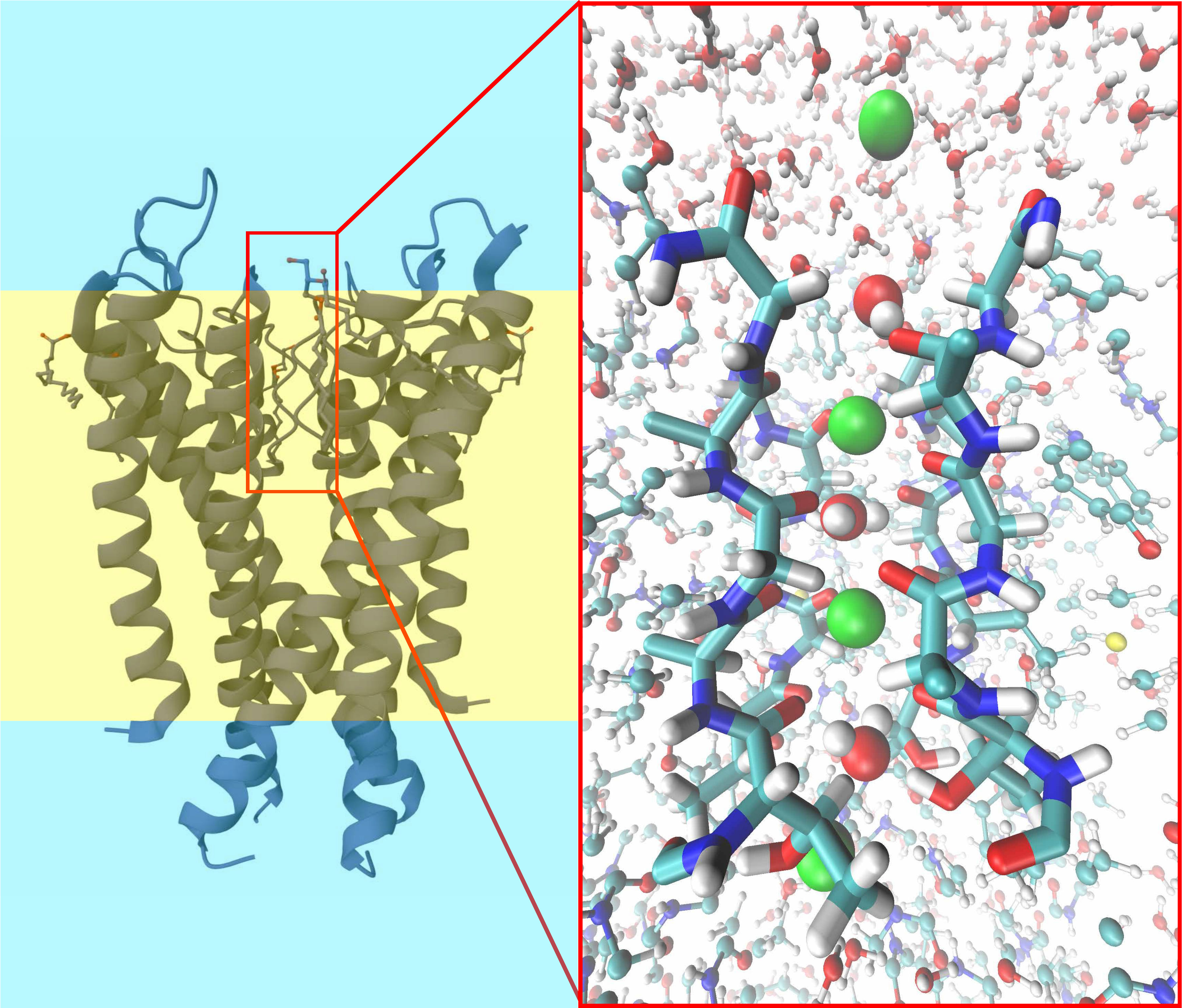}
\caption{The KcsA potassium ion channel (PDB ID: 1K4C) and its selectivity filter (SF).  Overall structure with the membrane region indicated in yellow is shown with a close-up view of the SF, which is formed by four identical TVGYG sequences arranged in tetrameric symmetry. The SF contains five layers of oxygen atoms that point into the SF and coordinate K$^+$ ions (green spheres) as they move single-file upwards through the SF from the intracellular to extracellular side. }
\label{fig:fig1}
\end{figure}
 
\section{Introduction}
\subsection{Background}
Throughout nature, a particular sequence of amino acids is repeatedly observed: TVGYG. This pattern signals the presence of a potassium ion channel, such as the archetypal KcsA channel found in bacteria. These channels are among the most fundamental components of biology. The TVGYG sequence encodes for the narrowest part of the channel—the selectivity filter (SF)—which is composed of four copies of this sequence (see Figure~\ref{fig:fig1}) and enables the transport of potassium ions across the cell membrane with remarkable speed and selectivity. \cite{doyleStructurePotassiumChannel1998,zhouChemistryIonCoordination2001} Understanding how these channels enable selective yet rapid potassium ion transport is critically important, as they carry the electrical signals central to a vast array of biological functions such as nerve signaling and muscle contraction.\cite{rouxIonChannelsIon2017}  Insights into their mechanism could also inspire the design of new advanced materials for ion-selective membranes. \cite{shiElectrolyteMembranesBiomimetic2020}

\paragraph{}

Despite their importance, we still do not understand many foundational questions about their behavior. In particular, there is still active debate about the transport mechanism: whether potassium ions move through the channel accompanied by water molecules (soft knock-on) or as a pure chain of ions (hard knock-on). The presence or absence of water co-transport through the SF has important implications. It affects the energetics of ion movement, influences cellular water homeostasis, and may be crucial for understanding the channel's exceptional K$^+$/Na$^+$ selectivity and deactivation mechanisms. 

Various early lines of evidence supported a soft knock on mechanism where water is co-transported with potassium.\cite{millerCouplingWaterIon1982,alcayagaStreamingPotentialMeasurements1989,aqvistIonPermeationMechanism2000,zhouChemistryIonCoordination2001,zhouMutantKcsAChannel2004,andoCoupledK+WaterFlux2005,iwamotoCountingIonWater2011} For example, streaming potential measurements, the most direct method of measurement,\cite{mironenkoPersistentQuestionPotassium2021} were used to correctly predict the dimensions of the SF\cite{millerCouplingWaterIon1982,alcayagaStreamingPotentialMeasurements1989} long before high-resolution crystal structures were available. Recent 2D-IR spectroscopy measurements also provide clear evidence of the presence of water molecules in the SF.\cite{kratochvilInstantaneousIonConfigurations2016,ryanProbingIonConfigurations2023}

However, there are also various experimental lines of evidence that support a hard knock on mechanism such as anomalous X-ray diffraction measurements.\cite{langanAnomalousXrayDiffraction2018} Ref~\cite{mironenkoPersistentQuestionPotassium2021} provides a comprehensive overview of the evidence for both mechanisms. 

\paragraph{}

Accurate molecular simulation is key to resolving this debate. Classical molecular dynamics (CMD) generally predicts a hard knock-on mechanism.\cite{furiniAtypicalMechanismConduction2009,kopferIonPermeationChannels2014,kopecDirectKnockonDesolvated2018}  However, CMD also consistently underestimates the rate of ion conductance by an order of magnitude or more.\cite{fowlerEnergeticsMultiIonConduction2013,jingThermodynamicsIonBinding2021,furiniCriticalAssessmentCommon2020,treptowIsoleucineGateBlocks2024} The magnitude of this underestimate implies that qualitative effects are missing in CMD simulations.

\subsection{Related work}

Neural network potentials (NNPs) are an exciting alternative to the classical force fields used in CMD.\cite{Behler2007,Behler2021,Kocer2022,batznerAdvancingMolecularSimulation2023,omranpourPerspectiveAtomisticSimulations2024} They use a neural network to predict the forces on the atoms in a given arrangement and are trained on quantum chemical calculations.

Recently, it was shown that when peptides were simulated with an NNP, they exhibited significantly more dynamic and flexible behavior compared to simulations using a classical force field. The better accuracy of the NNP was confirmed by comparison with spectroscopic measurements.\cite{unkeBiomolecularDynamicsMachinelearned2024}
This suggests that simulating the potassium ion channel with an NNP may reveal interesting new phenomena that could provide insight into the potassium ion transport mechanism.

However, the NNPs currently developed for biological simulation cannot handle potassium ions. Additionally, equivariant architectures, which provide state-of-the-art accuracy, are too slow to reach the time scales of interest for this process, and stability issues can emerge in long simulations.

Foundation or universal neural network potentials are another, even more exciting, recent development.\cite{chenUniversalGraphDeep2022,takamotoUniversalNeuralNetwork2022,dengCHGNetPretrainedUniversal2023,batatiaFoundationModelAtomistic2023,kovacsMACEOFF23TransferableMachine2023,merchantScalingDeepLearning2023,yangMatterSimDeepLearning2024a}  Universal NNPs are trained on large and diverse databases of quantum chemistry calculations and have shown evidence of strong generalization across multiple different chemistries. For example, MACE-MP-0, an equivariant universal NNP, has been used to simulate aqueous electrolyte solutions despite having been trained purely on DFT calculations of crystal structures. \cite{batatiaFoundationModelAtomistic2023} 

Orb-D3 is another recently released universal NNP. \cite{neumannOrbFastScalable2024} It is pretrained using a denoising objective then fine tuned on a large database\cite{jainCommentaryMaterialsProject2013,schmidtMachineLearningAssistedDeterminationGlobal2023} of crystal structure DFT calculations meaning it has a large exposure to ionic interactions. The training data uses the dispersion corrected Perdew-Burke-Ernzerhof (PBE) level of density functional theory (DFT). The model is not constrained to be equivariant or conservative and operates on raw distance vectors using data augmentation to learn equivariance approximately.\cite{pozdnyakovSmoothExactRotational2023,langerProbingEffectsBroken2024} This more efficient model architecture enables significantly faster simulations, while maintaining high accuracy. Orb-D3 also has remarkable transferability. For example, it is capable of simulating small molecules at 500 K despite having been trained on DFT optimization trajectories of crystalline materials performed at 0 Kelvin. \cite{neumannOrbFastScalable2024} 

\paragraph{}

Here, we simulate the SF with Orb-D3 showing it remains stable throughout the $\approx$ 1 ns long simulations. Additionally, a previously unreported hydrogen bond is observed that appears to enable water co-transport through the SF, along with an ion conductance comparable to experimental values, and carbonyl oxygen flipping at sites in the SF that have not been previously observed.

\section{Methods}
\begin{figure}[tbh]
\centering
\includegraphics[width=.5\columnwidth]{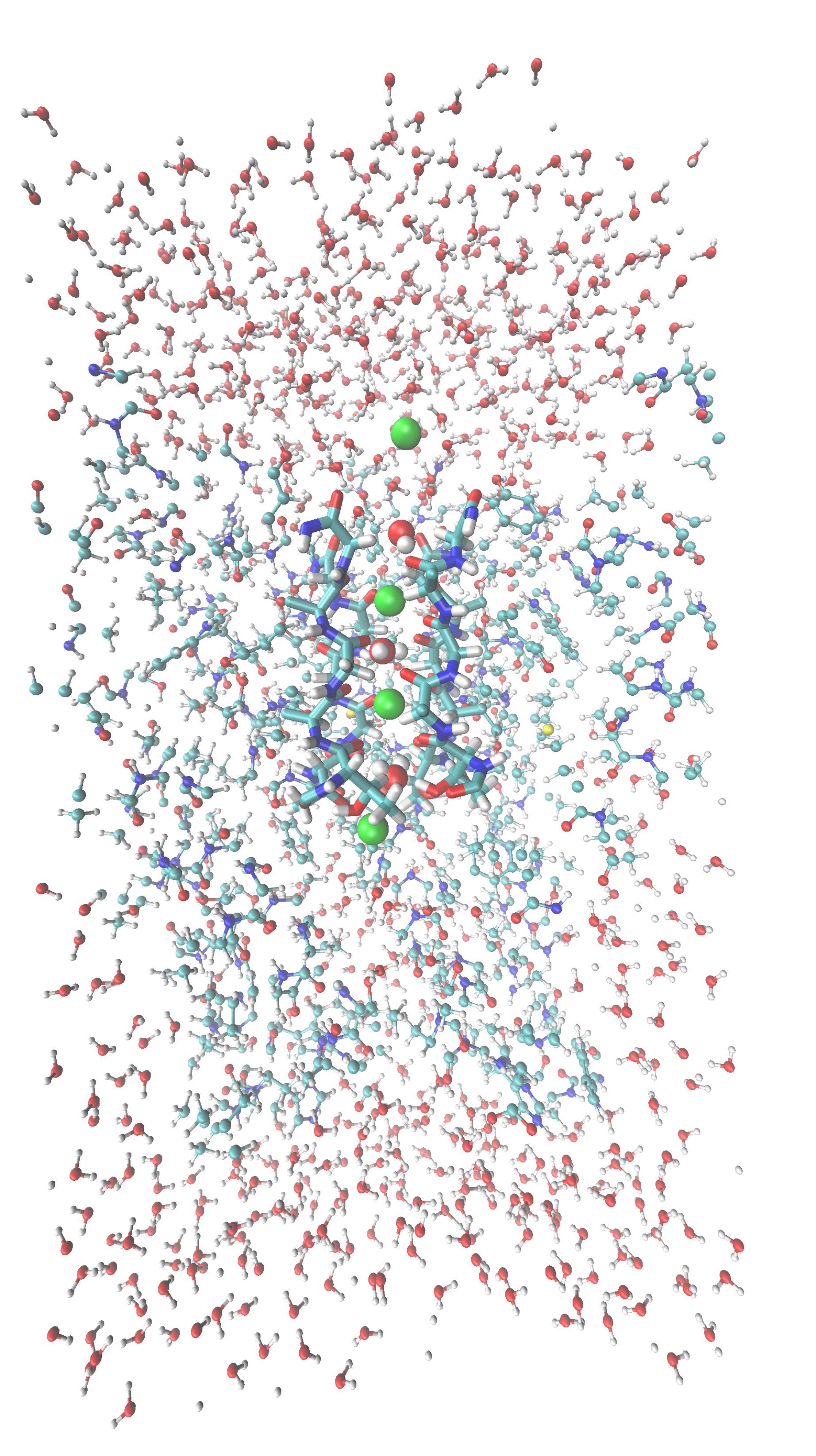}
\caption{Cross section of the full cylindrical system simulated. The SF and its contents are shown in bold. The outer 5 \AA\ of the cylinder are frozen in place to maintain the stability of the SF.}
\label{fig:Fullsystem}
\end{figure}

Orb-D3-v2 was used for all simulations presented here. Initial attempts to simulate the whole protein with Orb-D3 were unsuccessful this is unsurprising given that this is a membrane protein. Including surrounding lipids is not computationally feasible as it would require tens of thousands of atoms. Therefore, a 20 \AA\ radius cylinder of length 72.95 \AA\ was simulated. The SF lies along the $z$ axis of the cell and atoms more than 15 \AA\ away from the $z$-axis were frozen to stabilize the SF. The resulting simulation contains 8,450 atoms. A simulation cell with dimensions of 72.95$^3$\AA$^{3}$ was used (see Figure~\ref{fig:Fullsystem}). The system is connected in the z-direction via periodic boundary conditions. The initial structures were taken from an earlier simulation paper on this system.\cite{jingThermodynamicsIonBinding2021}  

The simulations are initialized with alternating potassium ions at the entrance S4 position and at the S2 position inside the channel, with water molecules at S1 and S3, as shown in Figure~\ref{fig:snapshots}. 

The ASE package\cite{hjorthlarsenAtomicSimulationEnvironment2017} was used to perform the simulation with a Langevin thermostat\cite{vanden-eijndenSecondorderIntegratorsLangevin2006}; a friction coefficient of 0.01 was chosen with a time step of 0.5 fs. The temperature was set to 300 K but the average temperature was actually 307 K. The extra 7 K is presumably attributable to a small amount of noise on the forces.

A large force was also applied to the ions  (0.1 eV\AA$^{-1}$) to accelerate their transport through the SF.  Simulations were run for $\approx$ 700 ps until a full conduction cycle had occurred. 

VMD was used for the visualization of the structures.\cite{spivakVMDPlatformInteractive2023}

\section{Results \& Discussion}
\subsection{Transport mechanism}
\begin{figure}[tbh]
\centering
\includegraphics[width=\columnwidth]{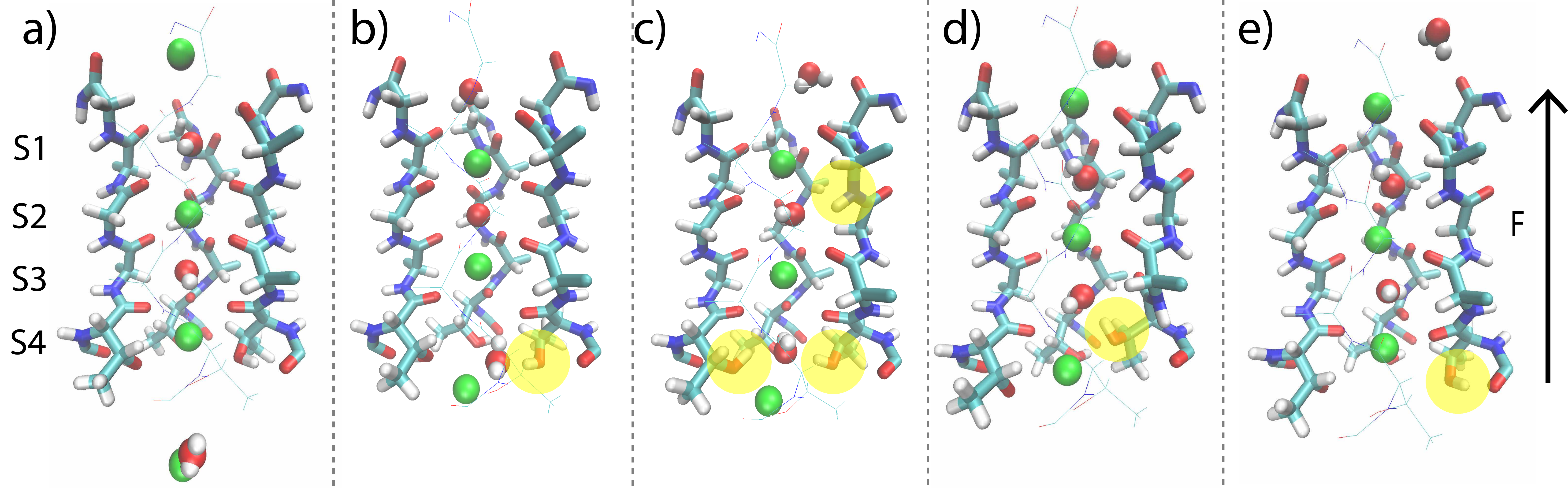}
\caption{Snapshots from a simulation trajectory showing rapid water co-transport through the SF of a potassium ion channel. The entrance site of the SF (S4) is surrounded by four hydroxyl side-groups and four backbone carbonyl oxygens of the threonine T75 residues. The following three sites (S3-S1) are each surrounded by eight carbonyl oxygens. (a) at $t=0$ ps the T75 hydroxyl groups at the entrance are hydrogen bonded to the carbonyl oxygen of the T74 residue further outside the channel creating a ring where the hydrogen bonds point away from the SF. (b) At $t= 20$ ps a hydrogen bond forms between the T75 hydroxyl side group and a water molecule outside the SF. (c) At $t=50$ ps the water molecule enters the SF and temporarily H-bonds to two T75 hydroxyl groups. One of the glycine (G77) residue carbonyls also flips. (d) At  $t=240$ ps the water hops to the S3 position and the potassium ion enters the S4 site. (e) At $t=250$ ps the T75 hydroxyl group returns to its starting configuration. Completing a full cycle. The protein on the front side of the SF is represented with thin lines to aid visualization.}
\label{fig:snapshots}
\end{figure}
\begin{figure}[tbh]
\centering
\includegraphics[width=.8\columnwidth]{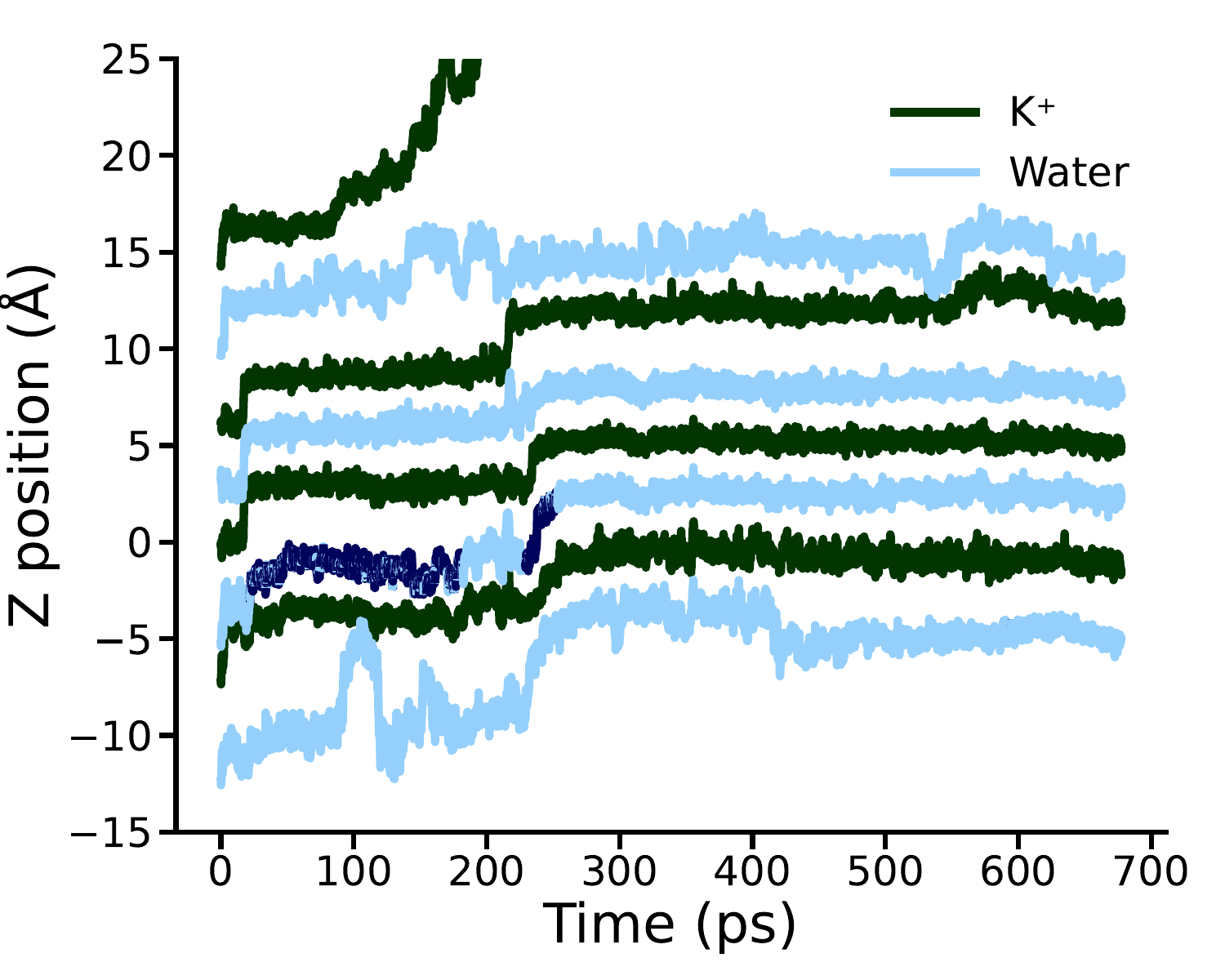}
\caption{Positions of potassium ions and water molecules through the SF as a function of time. Dark blue indicates when the water molecule is hydrogen bonded to one of the threonine residue side group hydroxyls at the entrance to the SF.}
\label{fig:z_positions_plot}
\end{figure}

With a large force applied to the potassium ions (greater than 0.1 eV/Å), this simulation appears stable as the potassium ions move through the selectivity filter. (SF)  However, with this force a hard knock-on mechanism of the potassium ions occurs. This is unsurprising as the water molecules do not have sufficient time to equilibrate and migrate into the SF. However, when a force of 0.1 eV/Å is applied, a surprising and previously unreported transport mechanism emerges, where a water molecule is co-transported through the SF with a soft knock-on mechanism as originally hypothesized. This trajectory is shown in Figure~\ref{fig:snapshots} and Figure~\ref{fig:z_positions_plot}.

This simulation shows a water molecule at S1 rapidly leave the SF.A correlated hopping event then occurs where the three remaining molecules in the channel simultaneously hop forward one position, leaving the S4 entrance position vacant.

Once the S4 site is vacant, the hydroxyl group on the side chain of the threonine T75 residue at the entrance of the channel reorients to hydrogen-bond with a water molecule near the entrance. This water molecule can then enter the S4 site. This contrasts with simulations initialized with a water molecule in the S4 position, where it rapidly hops out before this stabilizing hydrogen bond has time to form. 

Once the water molecule occupies the S4 site, the remaining three molecules hop forward, leaving a vacancy at the S3 site. The first water molecule then moves into S3 while still remaining hydrogen-bonded to the hydroxyl group at the entrance. This is enabled by the partial rotation of the backbone carbonyl oxygen on the threonine T75 residue. This hydrogen bond then breaks, and the SF returns to its original arrangement as initialized in the simulations, completing a full conduction cycle. 

\begin{figure}[tbh]
\centering
\includegraphics[width=.8\columnwidth]{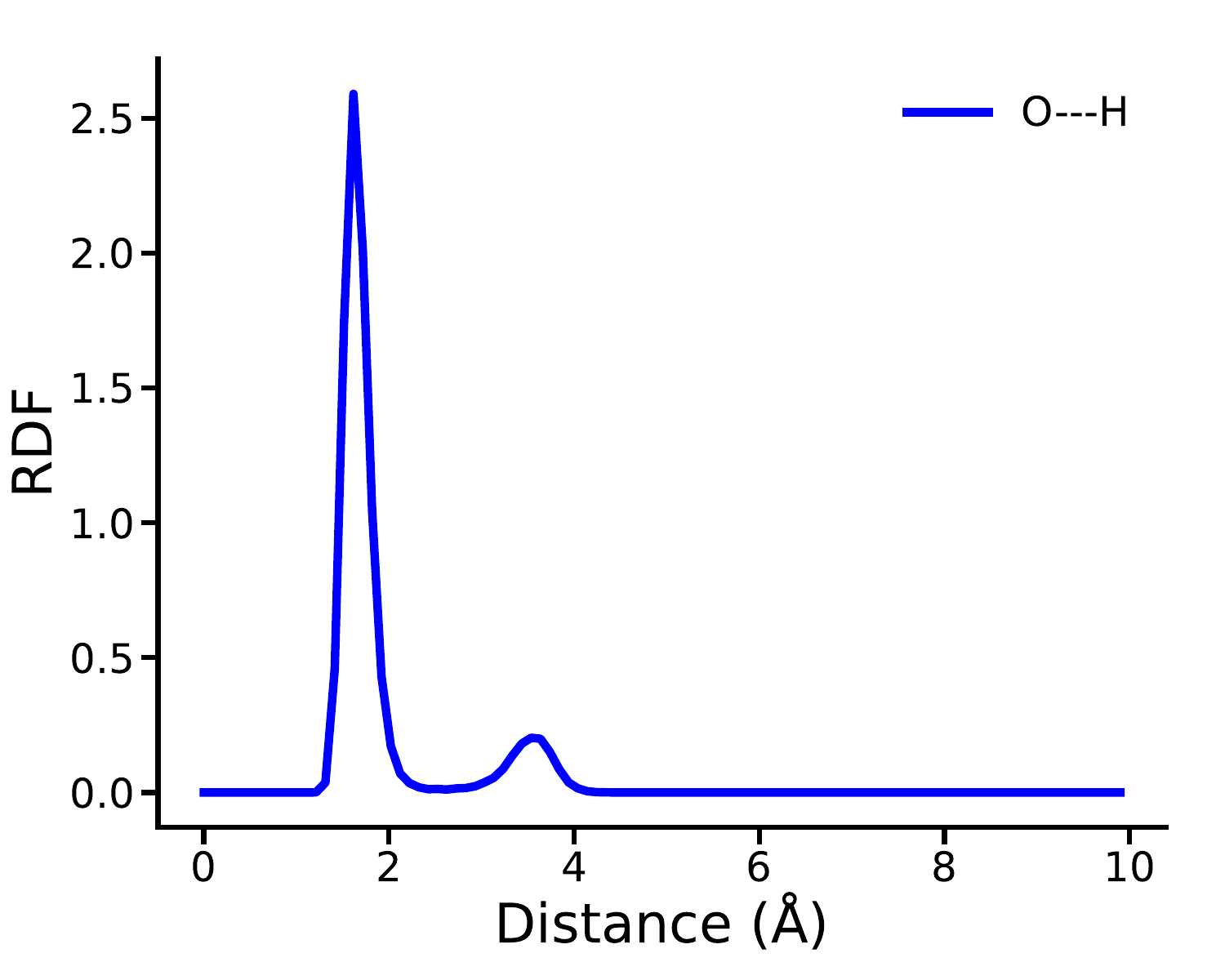}
\caption{Radial distribution function (RDF) of the O-H distance between the hydrogen of the hydroxyl side group of the threonine residue (T75) at the entrance to the SF and the neighboring carbonyl oxygen of the T74 residue outside the SF.}
\label{fig:average_rdf_plot}
\end{figure}

The radial distribution function (RDF) between the hydrogen of the hydroxyl side group of the threonine residue (T75) at the entrance to the SF and the neighboring carbonyl oxygen outside the SF is shown in Figure~\ref{fig:average_rdf_plot} showing two clear peaks. The larger distance peak corresponds to cases where the T75 hydroxyl side group is hydrogen bonded to water molecules in or near the SF. 

The formation of a hydrogen bond with the threonine side chain, which can reach even into the S3 site, appears to be key to enabling the movement of the water molecule into the SF. To the best of the author's knowledge, this mechanism has not been observed or suggested before. Its absence in CMD simulations, could explain why these simulations normally produce a hard knock-on transport mechanism with almost no water molecules in the SF, and significantly underestimate ion conductance by an order of magnitude or more.\cite{huiChargeScalingPotassium2024}

\subsection{Reproducibility}
Two additional identical simulations were run after the initial observation. One hard knock on and one soft knock on mechanism were observed. The two soft knock on mechanisms took a total of 0.25 and 0.05 ns for a full conduction cycle to occur, i.e., all atoms hopped twice. These simulations had no initial equilibration time. The initial simulations started with a potassium close to the entrance of the SF. Four more simulations were then run with an initial 60 ps pre-equilibration period where no force was applied on the potassium ions. Once a force was applied a soft knock on mechanism occurred in all four of these simulations and took 0.19, 0.43, 0.72 and 0.87 ns for a full conduction event to occur. In all six simulations hydrogen bonding of the T75 side group hydroxyl with water molecules at the entrance to the SF was observed although for varying durations. 

The average time of a full conduction cycle in the six simulations where it was observed was 0.4 $\pm$ 0.1 ns assuming a membrane thickness of 50 \AA\ this corresponds to a conductance of 80 $\pm$ 20 pS, which falls within the experimental range of 40-250 pS observed under symmetric conditions (20-800 mM K+, low voltage).\cite{lemasurierKcsAItsPotassium2001} 

However, it is important to note that the applied force of 0.1 eV/\AA\ corresponds to a trans membrane voltage of 5 V, which is much higher than physiological voltages which are typically 0 to 0.2 V. Additionally, the simulations are initiated with a potassium ion near the entrance to the SF so do not include the diffusion time of a potassium ion up to the entrance to the SF. The entrance to the SF also gradually dehydrates over the course of the simulation due to significant movements of surrounding protein. This may be caused by the  nonphysical restraints placed on outer region of the simulation to stabilise the SF. This prevents accessing longer time scales with the current system. 

\begin{figure*}
      \begin{subfigure}[]{0.49\textwidth}
        \centering
        \includegraphics[width=\textwidth]{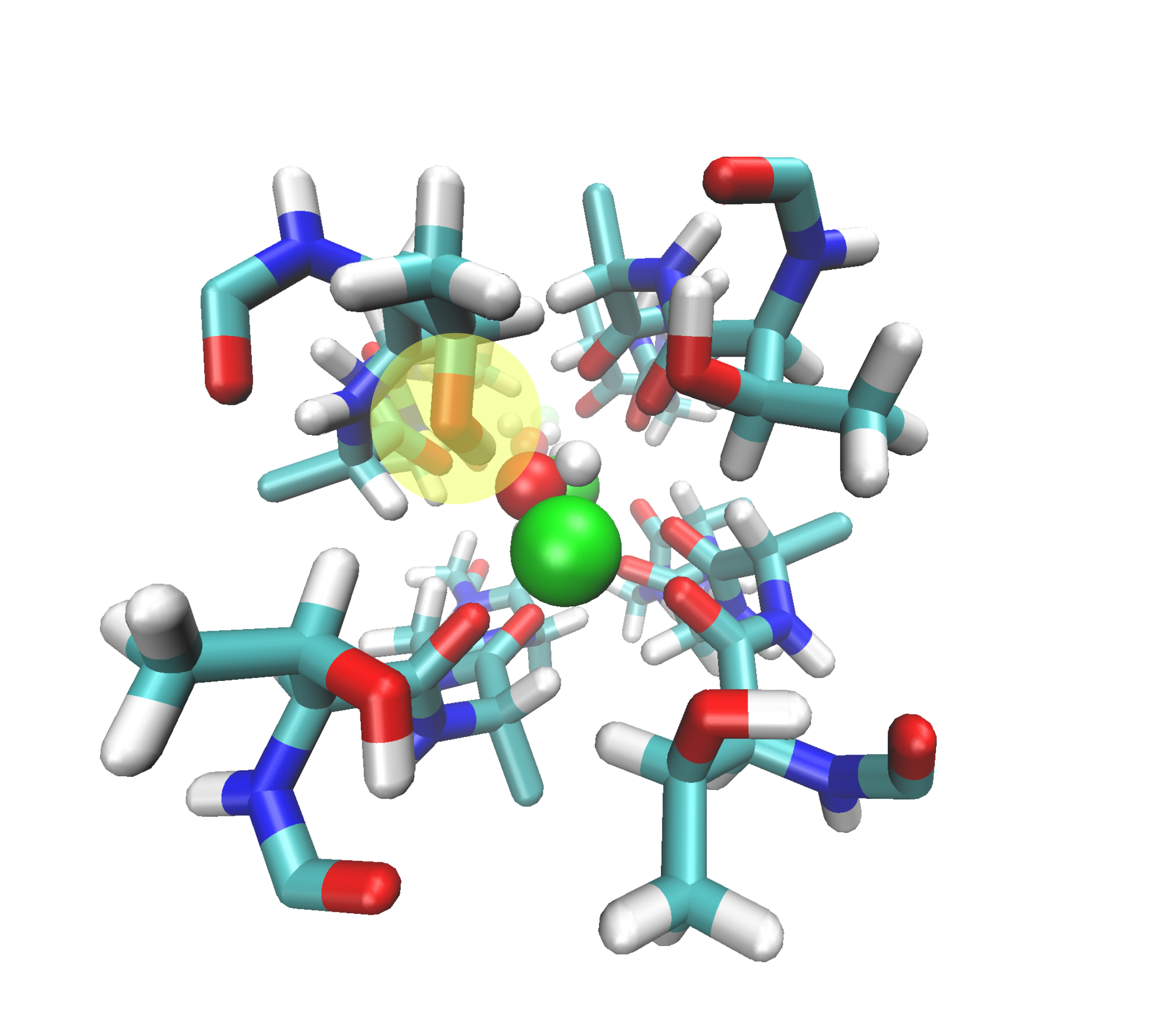}
        \caption{T75 Hydrogen bond. }\label{fig:Hbond}
    \end{subfigure}
        \begin{subfigure}[]{0.49\textwidth}
        \centering
        \includegraphics[width=\textwidth]{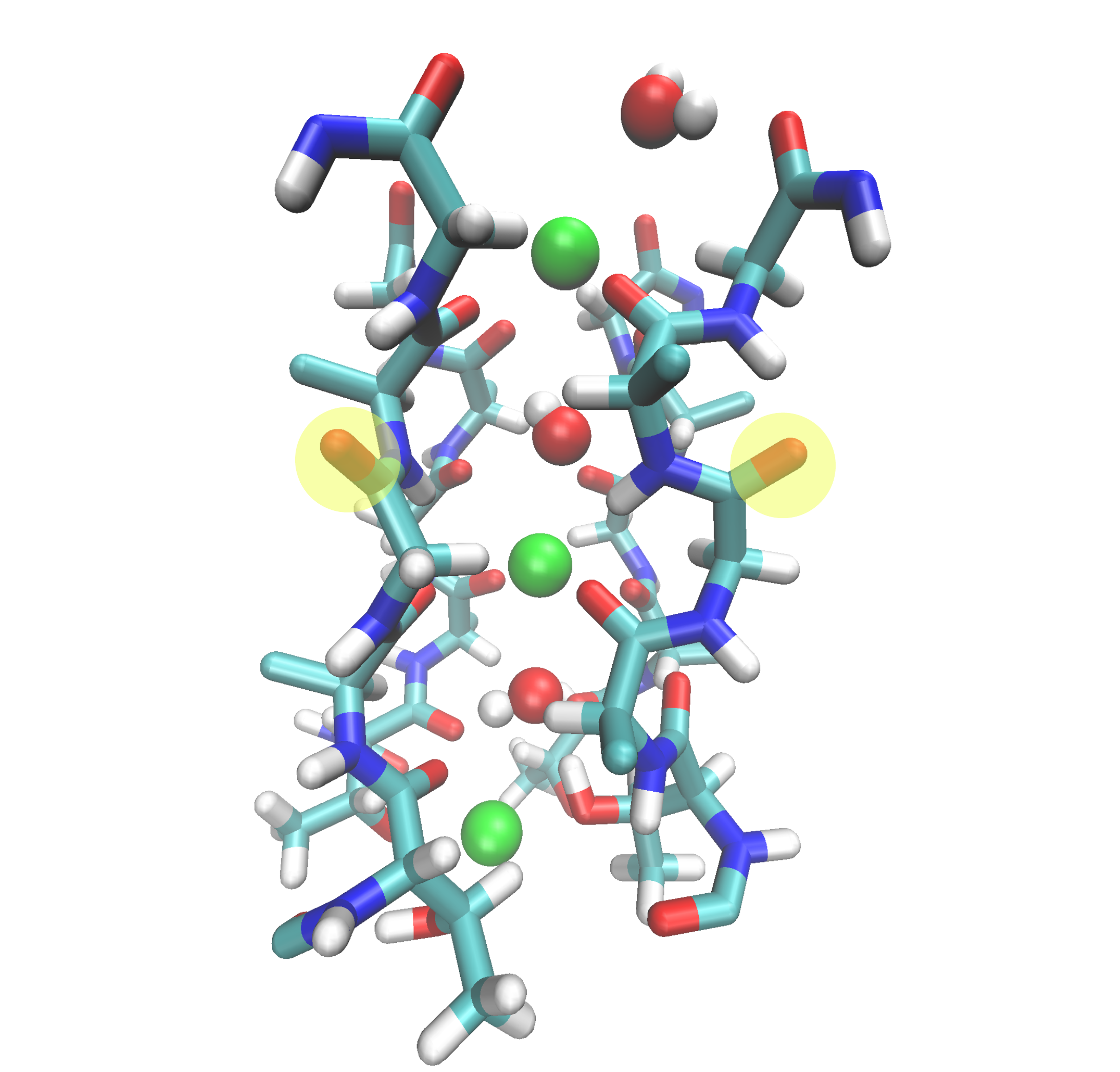}
        \caption{Flipped G77s carbonyls.}\label{fig:flippedG77}
    \end{subfigure}
    \caption{Example snapshots from the simulations. (a) The T75 hydroxyl group hydrogen bonded to a water molecule inside the SF.  (b) Two flipped G77 carbonyl oxygens.}
    \label{fig:closeups1}
\end{figure*}

\subsection{T75C mutations evidence}

If the T75 hydroxyl side-group plays an important role in the rapid conduction of potassium ions as observed here, we should expect the conductance to drop significantly if it is removed. Note that this is a counterintuitive prediction: the removal of a layer of a filter should make it conduct ions faster, not slower.

This prediction has, in fact, already been demonstrated.\cite{zhouMutantKcsAChannel2004} Mutating the threonine residue at the entrance of the SF to a cysteine almost perfectly removes this hydroxyl group without further altering the structure of the SF. This mutation significantly reduces the transport of potassium through the channel, bringing it down to a level similar to that of rubidium ions. In contrast, rubidium’s conductance through the channel only increases slightly from this mutation, as would be expected from removing a layer from a filter.\cite{zhouMutantKcsAChannel2004} This provides strong evidence that the hydroxyl group plays a key causative role in the rapid, specific transport of potassium ions through the channel. 

The reduced density at the S4 site when the hydroxyl group is removed in experiments indicates that the potassium ion is destabilized there, disrupting the balanced free energies of the two states believed to be key to enabling rapid conductance.\cite{morais-cabralEnergeticOptimizationIon2001}

\subsection{Additional observations}
\textbf{Carbonyl flipping of G77 residues.}
During our simulations, we also observe a flipping of two of the G77 residues induced by the presence of a water molecule—a phenomenon not previously observed in classical MD simulations or experiment to the best of the author's knowledge. This is shown in Figure~\ref{fig:snapshots}c, as well as in a close-up view in Figure~\ref{fig:flippedG77}. This flipping may explain why the S2 site has a low ion occupancy at low concentrations of potassium.\cite{boiteuxSelectivityFilterIon2020} A similar effect has been observed with rubidium.\cite{morais-cabralEnergeticOptimizationIon2001}

\textbf{Carbonyl flipping of two V76 residues.} Carbonyl flipping of two of the V76 residues is also observed in some simulations. In both cases, the flipping appears to be induced by the presence of a water molecule, which creates a binding site for the water with two hydrogen bond donors and two acceptors. 

The flipping of V76 residues has been observed in classical molecular dynamics (CMD) simulations and has also been hypothesized to relate to the mysterious c-type inactivation of potassium ion channels where the SF gradually loses conductivity due to conformational changes in the SF.\cite{bernecheGateSelectivityFilter2005}  It is known that the SF can switch to a closed conformation where two water molecules occupy positions behind the G77 and V76 residues.\cite{zhouChemistryIonCoordination2001} These simulations suggest the possibility that the water molecules could be carried to these positions by this flipping. 

The flipping of these residues provide hydrogen bond donors to water molecules in the SF, which may explain why pure water is experimentally observed to pass through the SF very quickly.\cite{saparovDiffusionLimitWater2004,hoomannFilterGateClosure2013}

\textbf{Carbonyl flipping of T74 residue.} 
In some simulations when the T75 hydroxyl group hydrogen bonds to a water in the SF, the carbonyl of the T74 residue it is normally hydrogen bonded to can also flip, preventing the hydroxyl group from returning to its original orientation. When this occurs the T75 hydroxyl hydrogen bonds to a water in the SF or the first carbonyl oxygen inside the SF.

\textbf{Potassium only SF.}
Intriguingly, when the simulation is initialized with four potassium ions and no water in the SF, the soft knock-on mechanism is not observed, and a hard knock-on mechanism occurs. In fact, the strong positive charge in the SF appears to cause the T75 hydroxyl side-group to hydrogen-bond to a water molecule outside the SF, allowing a potassium ion to enter the SF without a water. Therefore, a hard knock on mechanism may  still be possible under some conditions.  

\textbf{Sodium ions enter the SF.}
Initial simulations with sodium ions instead of potassium show these ions can enter the SF without water, but do not move out quickly. This indicates that the standard selectivity mechanism, i.e., the free energy cost of dehydrating the sodium is too large, may be an oversimplification and more complex mechanisms may be involved.\cite{mitaConductanceSelectivityNa2021}

One possibility is that sodium may trigger conformation changes in the SF to a closed state enabling the SF to act as a kind of Maxwell’s demon, selectively closing when it detects sodium ions. This could provide a much stronger selectivity mechanism. 

\textbf{H25 deprotonation.} Towards the end of the simulation, the deprotonation of the histidine (H25) residues on the intracellular side of the channel occurs. This is consistent with the fact that this channel is known to be gated with a low pH to allow the conduction of potassium ions. There is evidence this is controlled by the protonation state of this histidine H25 residues.\cite{swensonChannelsCloseMore1981} This suggests that it may be possible to examine the pH-dependent gating of the channel using NNPs.

\subsection{Bulk electrolyte}
\begin{figure*}
    \begin{subfigure}[]{0.49\textwidth}
        \centering
        \includegraphics[width=\textwidth]{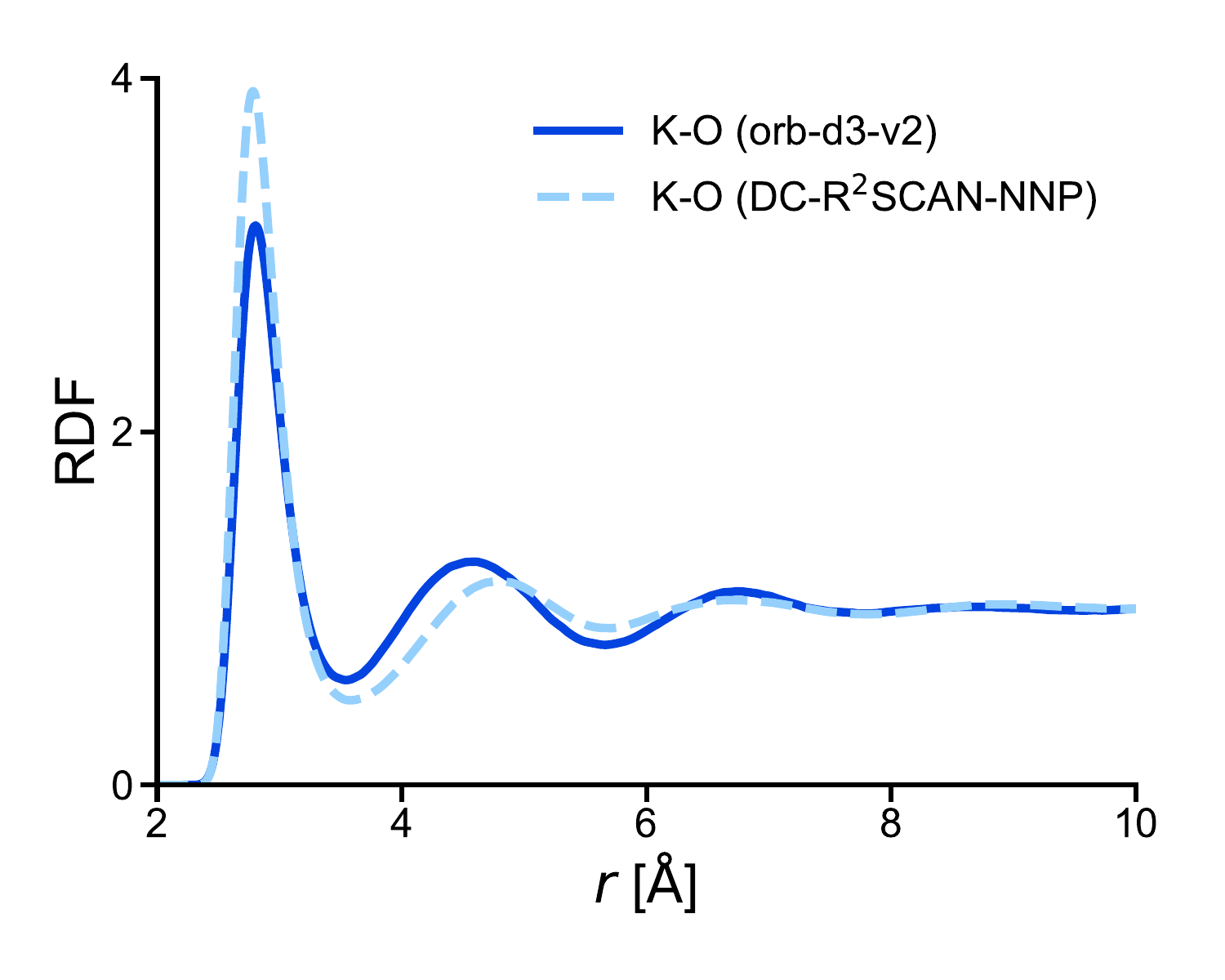}
        \caption{Potassium-oxygen RDFs}\label{fig:gofrKO}
    \end{subfigure}
    \begin{subfigure}[]{0.49\textwidth}
        \includegraphics[width=\textwidth]{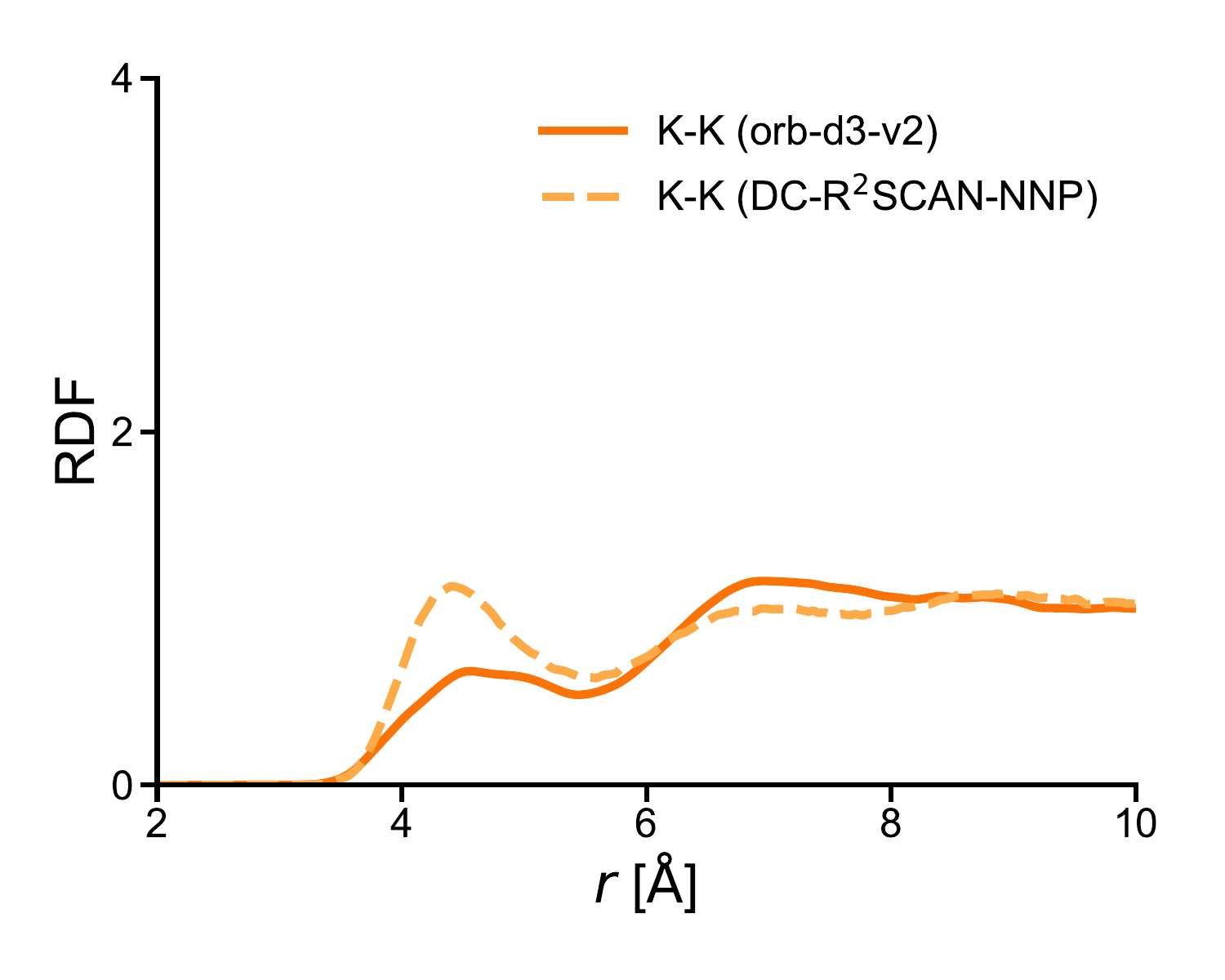}
        \caption{Potassium-potassium RDFs}\label{fig:gofrKK}
    \end{subfigure}
    \caption{Comparison of RDFs computed with Orb-D3 compared with a custom trained NNP using higher level DFT data. (Density corrected DFT SCAN) The simulation is of a 2.4 M KCl solution, taken from Ref.~\cite{zhangScalableMolecularSimulation2023}.}
    \label{fig:RDFs}
\end{figure*}

One objection to the use of the Orb-D3 model could be that it is trained on both neutral and charged potassium ions and hence might not reliably simulate potassium ions, as it cannot distinguish their charge state. To test this, a 2.4 M KCl aqueous solution was simulated and compared the K–O and K-K RDFs with another recently reported neural network potential simulation,\cite{zhangScalableMolecularSimulation2023} which was custom-trained on higher-quality and directly relevant training data. As shown in Figure~\ref{fig:RDFs}, the RDFs show good agreement between the two models. The K–O RDFs are also in good agreement with direct first-principles simulations of the potassium ion in water.\cite{duignanQuantifyingHydrationStructure2020} There are some quantitative discrepancies, likely attributable to the PBE-D3 level of theory used by Orb-D3.  s

\section{Future Directions}
There are a number of important next steps for this work to test the observations made here. Firstly, adding directly relevant training data, ideally with a higher level of DFT theory will be important. 

Secondly, the simulations reported in this paper only model part of the potassium ion channel cutting out portions of the surrounding protein. There may be important allosteric behavior that requires a larger region of the protein to be included and treated dynamically. Hybrid approaches with classical MD could enable this or scaling the simulation to many GPUs. 

Thirdly, the large force applied to the potassium ions (0.1 eV\AA$^{-1}$) is much larger than physiological conditions and may induce distortions in the transport mechanism. Therefore,  longer timescales with a more realistic applied force should be used. This will be important for exploring additional effects such as c-type inactivation, flickering conductance, ion selectivity, and the transition to the closed state.  In these simulations the time scale we can access is limited by significant dynamic flexibility of the protein on the intracellular side of the channel, which results in gradual dehydration of the entrance to the SF and the formation of a gap between the dynamic region and the surrounding frozen atoms.  Resolving this issue should enable the study of longer time scales with smaller forces. 

Finally, more rigorous enhanced sampling techniques should be implemented to probe the potential of mean force associated with the movement of potassium ions through the channel and to precisely compute conductivities and other properties.

\section{Conclusion}
A universal neural network potential (Orb-D3), trained using DFT calculations on crystal structures, was used to simulate the selectivity filter of the KcsA potassium ion channel. Not only were stable simulations of ion conductance possible, but new features of the conduction mechanism were observed. While these initial observations need to be independently verified, if true this would be a remarkable demonstration of the transferability of universal neural network potentials.

These simulations indicate that the threonine T75 hydroxyl side group at the entrance to the SF plays a crucial role in stabilizing water molecules in the selection filter via a hydrogen-bond, enabling a soft knock-on transport mechanism. We also observed the flipping of the T74, V76, G77 carbonyl oxygen's induced by the presence of water molecules in the SF and an experimentally plausible conductivity of 80 $\pm$ 20 pS.

These simulations provide a clear demonstration of the importance of the flexible, dynamic nature of proteins.\cite{ballHowLifeWorks2024}. Until now, we have lacked adequate tools to reliably study these behaviors due to the extremely short timescales they occur on as well as their sensitivity to small errors in the potential energy surface. This work demonstrates the potential of universal NNPs to serve as a powerful new tool for overcoming this problem.\cite{duignanPotentialNeuralNetwork2024,defabritiisMachineLearningPotentials2024} It may now be possible to simulate a vast range of important biological, industrial, and geological processes accurately at the molecular scale for the first time. This will provide a wealth of new insights into many critically important phenomena. 

\section{Acknowledgment}
I would like to acknowledge Susan Rempe and Mark Stevens for helpful discussions and Benedict Irwin and Mark Neumann for feedback on the manuscript. This research was undertaken with the assistance of resources and services from the National Computational Infrastructure (NCI), which is supported by the Australian Government.

\section{Code availability}
Initial structures, simulation input files and analysis scripts are available at: \newline \href{https://github.com/timduignan/KcsA-SF_Orb-d3-v2}{github.com/timduignan/KcsA-SF\_Orb-d3-v2}
A video of the main trajectory can be found at: 
\href{https://www.youtube.com/shorts/LGd8R5SCR9U}{https://www.youtube.com/shorts/LGd8R5SCR9U}.
\newpage
\printbibliography

\end{document}